\documentclass[conference,compsoc]{IEEEtran}

\usepackage[T1]{fontenc}

\usepackage{graphicx}
\usepackage{xcolor}
\usepackage{hyperref}
\usepackage{amsmath}

\ifCLASSOPTIONcompsoc
  \usepackage[nocompress]{cite}
\else
  \usepackage{cite}
\fi

\ifCLASSINFOpdf
\else
\fi

\hypersetup{
    colorlinks=true,
    citecolor=magenta,
    linkcolor=purple,
    urlcolor=blue
}

\begin{document}

\title{Topology-Dependent Enhancement of Entanglement Extraction in Repeater Graph States}

\author{
    \IEEEauthorblockN{
        Poramat Chianvichai\IEEEauthorrefmark{1},
        Poramet Pathumsoot\IEEEauthorrefmark{2},
        Naphan Benchasattabuse\IEEEauthorrefmark{2},\\
        Michal Hajdu\v{s}ek\IEEEauthorrefmark{2},  
        Rodney Van Meter\IEEEauthorrefmark{3},
        Sujin Suwanna\IEEEauthorrefmark{1} 
    }\\
    \IEEEauthorblockA{
        \IEEEauthorrefmark{1}\textit{Optical and Quantum Physics Laboratory, Department of Physics, Faculty of Science,}\\\textit{Mahidol University, Bangkok 10400, Thailand}}
    
    \IEEEauthorblockA{
        \IEEEauthorrefmark{2}
        \textit{Graduate School of Media and Governance, Keio University Shonan Fujisawa Campus, Kanagawa 252-0882, Japan}
    }
    \IEEEauthorblockA{
        \IEEEauthorrefmark{3}\textit{Faculty of Environment and Information Studies, Keio University Shonan Fujisawa Campus, Kanagawa 252-0882, Japan}
    }
    
    e-mails: poramat.cha@student.mahidol.edu,\,\{poramet, whit3z, michal, rdv\}@sfc.wide.ad.jp,\, sujin.suw@mahidol.ac.th    
}

\maketitle

\begin{abstract}
Quantum repeaters are essential for establishing long-distance quantum communication to overcome the exponential decay of entanglement due to photon loss. Traditional repeater architectures rely on physical quantum memory, which introduces decoherence and poses significant practical implementation challenges. The repeater graph state (RGS) architecture offers a promising memory-less alternative that is inherently resilient to photon losses. A key challenge in implementing RGS lies in the requirement for highly efficient graph state generators and complex qubit measurement. In this work, we aim to investigate the strategies for extracting the maximum number of Bell pairs from the RGS structure via its qubit connection to resolve the well-known bottleneck problem of RGS in which only a single Bell pair can be extracted from a complete bipartite graph state. From simulations, we observe that the maximum number of extracted Bell pairs depends on its connection topology, where the Bell-pair yield tends to be maximal at low to moderate edge densities. As the number of network hops increases, the RGS must be equipped with higher inner-qubit connectivity to maintain a sufficient yield of extractable Bell pairs. Thus, the expected resource requirement shifts toward the use of RGSs with higher inner-qubit connectivity.
\end{abstract}

\IEEEpeerreviewmaketitle

\section{Introduction}
Quantum repeaters are essential and widely recognized as key components for the establishment of a large-scale quantum network \cite{RevModPhys.95.045006, Ruihong_2019, PhysRevLett.81.5932}. Quantum communication channels, whether they are deployed through free-space links or optical fibers, suffer from photon loss due to signal attenuation \cite{PhysRevA.106.042437, PhysRevA.106.L010401, Bäuml_2017}. The probability of success drops exponentially with distance, severely limiting the range over which the entanglement can be successfully distributed. Quantum repeaters circumvent this limitation by dividing the long communication line into shorter segments. However, traditional architectures fundamentally rely on physical quantum memory \cite{7010905, PhysRevA.75.032310}, a challenging requirement that itself can act as a source of decoherence. As a promising memory-less alternative, the all-photonic quantum repeater architecture utilizes entangled photons in a graph topology as primary information carriers \cite{Azuma2015-rm, PhysRevX.10.021071}.

While resilient to errors and photon loss, existing all-photonic schemes suffer from substantial resource overhead, often consuming many entangled states to extract a single Bell pair \cite{benchasattabuse2024architecture}. To address this problem, recent work explored extracting multiple Bell pairs from a single photonic resource state to significantly improve the throughput \cite{PhysRevLett.134.190801}. Building on the structural framework proposed by Benchasattabuse \textit{et al}.\cite{benchasattabuse2024architecture}, this work focuses on the repeater graph state (RGS) to investigate measurement strategies enabling multi-Bell-pair extraction. Through simulation, we determine the maximum number of Bell pairs extractable from various RGS configurations, providing direct insights into the efficiency, scalability, and practical feasibility of multi-Bell-pair all-photonic networks.

\section{Graph States}
We briefly introduce the concept of graph states. This fundamental concept is essential for understanding the behavior of the all-photonic repeater architecture and how its state is represented within the stabilizer formalism \cite{poulin2005stabilizer}. A more comprehensive graph-state theory and its application in quantum information can be found in Refs.\cite{hein2004multiparty, hein2006entanglement}.

\subsection{Graph State Definition}
\begin{figure}[h!]
    \centering
    \includegraphics[width=0.45\textwidth]{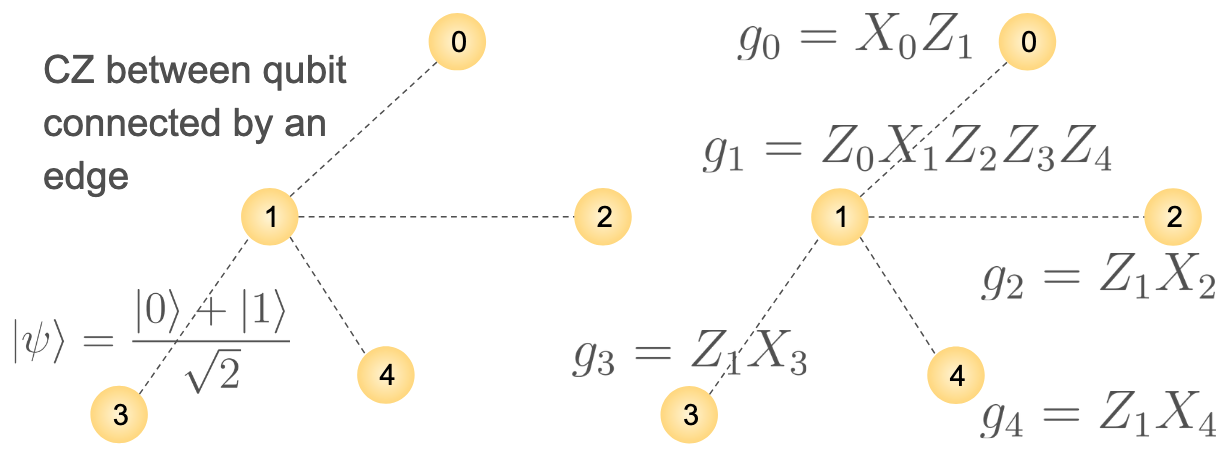}
    \caption{(left): Graph state representation using quantum state, each node state is $|+\rangle$ and connect with $CZ$. (right): Graph state representation using stabilizer generator formalism.}
    \label{fig:graph_state}
\end{figure}

Graph states form a family of quantum states described by a mathematical graph $G(V,E)$, where $V$ denotes the set of vertices, and $E$ the set of edges connecting the vertices. In this formulation, each vertex corresponds to a qubit prepared in the state $|+\rangle$, and each edge indicates an interaction between two qubits through a controlled-phase gate
\begin{equation}
    CZ = |0\rangle\langle 0|\otimes I +|1\rangle\langle 1|\otimes Z.
\end{equation}
Consequently, a graph state $|G(V,E)\rangle$ constructed from $G(V,E)$ can be expressed as 
\begin{equation}
    |G(V,E)\rangle = \prod_{u,v \in E} CZ(u, v)|+\rangle^{\otimes |V|},
    \label{eq:graph_state_eq}
\end{equation}
where $u$ and $v$ are vertices in the set $V$ forming an edge in the set $E$. It can be illustrated in Fig. \ref{fig:graph_state} (left).

By definition, the graph states form a subset of stabilizer states \cite{gottesman1997stabilizercodesquantumerror}. Therefore, each graph state can be represented in the stabilizer-generator form,
\begin{equation}
    g_u = X_u \prod_{v\in N_{u}} Z_v,
\end{equation}
with each vertex $u \in V$, and $N_u$ is the set of all neighbor vertices of $u$, as illustrated in Fig. \ref{fig:graph_state} (right).

\subsection{Measurement on Graph States}
Performing Pauli measurements can be effectively implemented by simple graph transformation rules, with the only requirement being local modifications that depend on the measurement outcome. For instance, measuring a single qubit in a specific Pauli basis ($X$, $Y$, or $Z$) on a graph state $|G(V,E)\rangle$ collapses the state. After accounting for measurement-outcome-dependent local corrections (byproduct operators), the remaining $N-1$ qubit states typically form another graph state on the modified graph $|G\rangle ^{\dagger}$. Let $\tau_{i}\{G\}$ denote the locally complement neighborhood of $i$. Then, the specific graph-theoretic transformation rules for the Pauli operators are as follows.

Pauli-$Z$: delete vertex $i$ from $|G(V,E)\rangle$
\begin{equation}
    G \rightarrow G - i
\end{equation}

Pauli-$Y$: Locally complement the neighborhood of $i$ and remove vertex $i$,
\begin{equation}
    G \rightarrow \tau_{i}\{G\} - i.
\end{equation}

Pauli-$X$: For any qubit $b$ neighboring $i$ ($b \in N_i$), locally complement $b$, apply the rule for the Pauli Y, then locally complement $b$ again,
\begin{equation}
    G \rightarrow \tau_{b}(\tau_{i} (\tau_{b}\{G\}) - i)
\end{equation}

\begin{figure}[h!]
    \centering
    \includegraphics[width=0.45\textwidth]{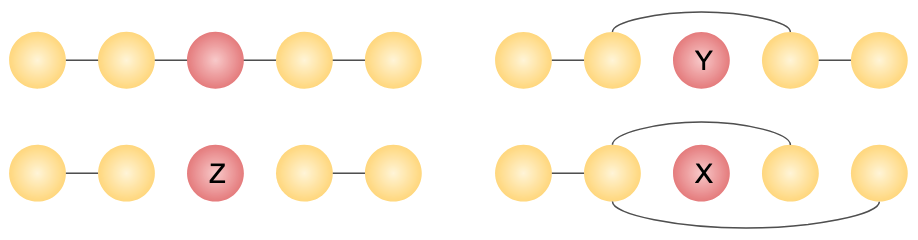}
    \caption{Graph transformations after performing Pauli $X$, $Y$, and $Z$ measurements on the middle qubit (red ball) in the linear graph state is that the measured qubit becomes disconnected from the graph, with all of its connecting edges removed. Moreover, the selected measurement basis determines how the remaining edges in the graph are updated.}
    \label{fig:measurement_on_graph}
\end{figure}

Graph transformations after applying different Pauli measurements on each qubit are illustrated in Fig. \ref{fig:measurement_on_graph}.

\section{Repeater Graph States}
The repeater graph states (RGS) lay the foundation of the multiple-qubit photonic states essential for realizing the all-photonic quantum repeater architecture. Key advantages include overcoming initial resource limitations through deterministic RGS generation schemes, utilizing a single quantum emitter, which would replace exponentially inefficient probabilistic methods. Moreover, generalized RGS states \cite{PhysRevLett.134.190801}, which borrow principles from capacity-achieving erasure codes to enable the simultaneous establishment of multiple Bell pairs, significantly improve scalability and performance \cite{Zhang:22}.

\begin{figure}[h!]
    \centering
    \includegraphics[width=0.45\textwidth]{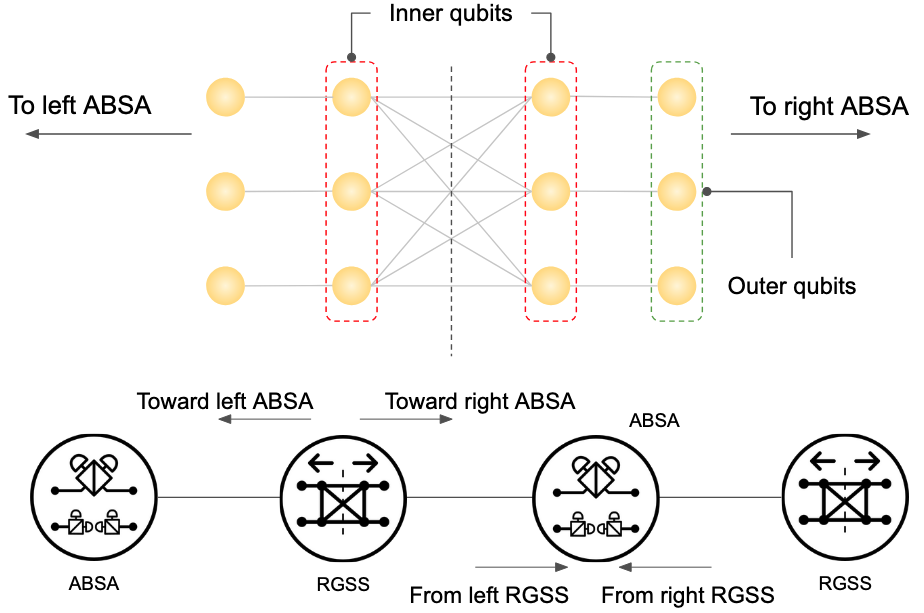}
    \caption{An RGS is depicted with three arms and six inner qubits forming complete bipartite graph. To create an entangled connection between the end nodes, each source node prepares an RGS and transmits half of its photons (yellow balls) to the left measurement node and the remaining half to the right measurement node. Full communication procedure using ABSA and RGSS shows in below diagram, node illustration from \cite{QuISP_github}}
    \label{fig:repeater_graph_state}
\end{figure}

Each RGS characteristically has $m$ arms in its structure, and it contains two types of qubits; namely, $2m$ inner qubits and $2m$ outer qubits. The inner qubits are arranged in a complete bipartite graph, forming the core of the RGS. Each inner qubit is then connected to a corresponding outer qubit as illustrated in Fig.~\ref{fig:repeater_graph_state}. 

In a communication scheme, there are two types of nodes: the RGS source nodes (RGSS) and the advanced Bell state analyzer nodes (ABSA) \cite{Benchasattabuse:2023qxi}. They operate analogously to conventional memory-assisted repeater nodes and standard Bell state analyzer (BSA) nodes, respectively. This architecture is engineered specifically to provide protection against the dominant challenge of long-distance quantum communication—photon loss—by embedding Quantum Error Correction (QEC) through tree-graph-encoded logical qubits \cite{PhysRevLett.97.120501}, thus circumventing the decoherence and storage limitations of matter-based quantum repeaters \cite{patil2024improveddesignallphotonicquantum}.

\section{Modeling Approach}
\begin{figure}[h!]
    \centering
    \includegraphics[width=0.47\textwidth]{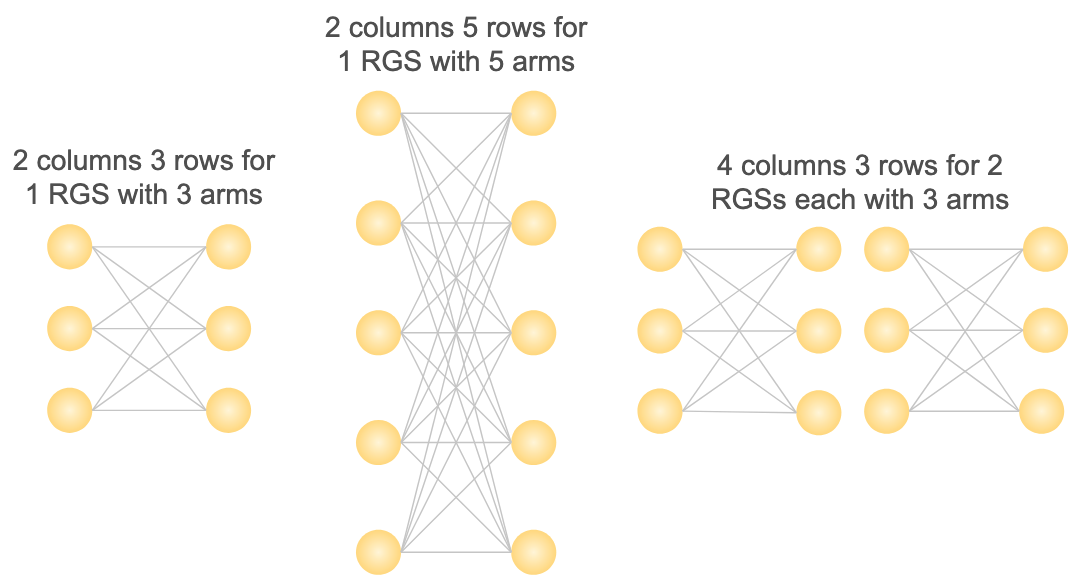}
    \caption{Topology of RGS: (left) a single RGS with three arms, (middle) a single RGS with five arms, and (right) two RGSs, each with three arms. }
    \label{fig:qubit_row_rgs}
\end{figure}

Our simulation begins by determining the number of RGSs required for the network by specifying the number of node columns, where every two columns correspond to a single RGS, as illustrated in Fig.~\ref{fig:qubit_row_rgs}. Increasing the number of columns directly scales the number of RGSs involved in the entanglement distribution process. 

\begin{figure}[h!]
    \centering
    \includegraphics[width=0.45\textwidth]{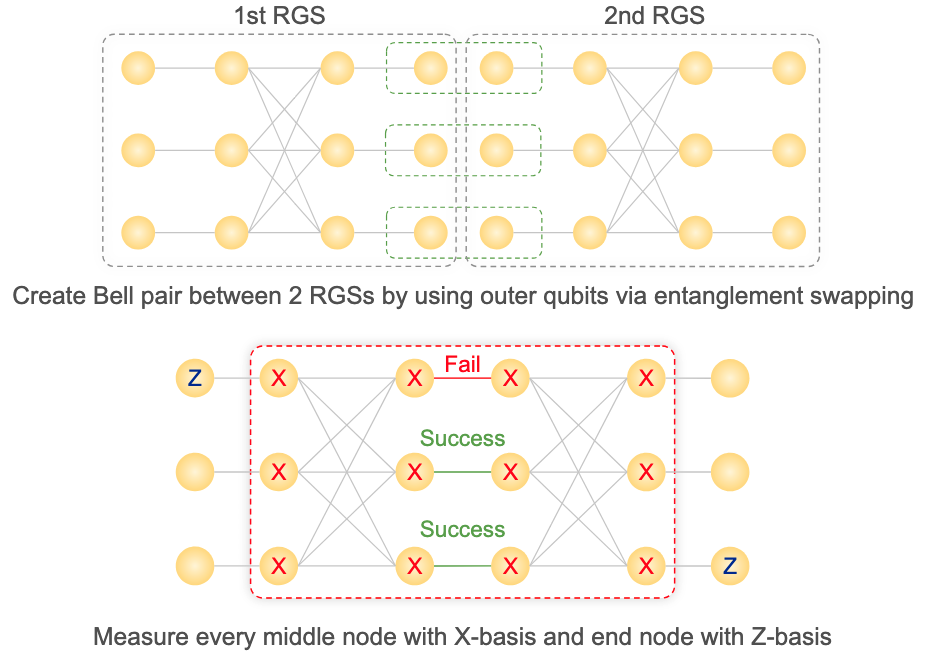}
    \caption{After connecting between inner qubits of different RGSs, measurments follow the procedure for extracting Bell pairs from the RGS network. Successful results create Bell pairs for communication resources.}
    \label{fig:rgs_measurement}
\end{figure}

The Bell-state measurement (BSM) rate serves as the key factor governing the probabilistic connections between adjacent RGSs. After generating graph state, we apply the stabilizer formalism to transform and analyze the resulting structure. The Bell pairs are then extracted by performing the $X$-basis measurements on all intermediate nodes and the $Z$-basis measurements on the end nodes of RGS as illustrated in Fig.\ref{fig:rgs_measurement}. Finally, the extracted stabilizers are compared with the ideal stabilizer generators corresponding to a valid Bell pair for the targeted RGS configuration, as listed in table \ref{tab:bell_stabilizer_results}.

\renewcommand{\arraystretch}{1.1} 

\begin{table}[]
\centering
\caption{List of stabilizers for extracting various Bell pairs from different RGS configurations.}
\label{tab:bell_stabilizer_results}
\begin{tabular}{|c|c|c|}
\hline
\textbf{1 RGS with 2 arms} & \textbf{1 RGS with 3 arms} & \textbf{2 RGSs with 3 arms} \\ \hline
\begin{tabular}[c]{@{}c@{}}
$X_0 Z_2$ \\ $Z_0 X_2$ \\ $X_0 Z_3$ \\ $Z_0 X_3$ \\ 
$X_1 Z_2$ \\ $Z_1 X_2$ \\ $X_1 Z_3$ \\ $Z_1 X_3$
\end{tabular} & 
\begin{tabular}[c]{@{}c@{}}
$X_0 Z_3$ \\ $Z_0 X_3$ \\ $X_0 Z_4$ \\ $Z_0 X_4$ \\ 
$X_0 Z_5$ \\ $Z_0 X_5$ \\ $X_1 Z_3$ \\ $Z_1 X_3$ \\ 
$X_1 Z_4$ \\ $Z_1 X_4$ \\ $X_1 Z_5$ \\ $Z_1 X_5$ \\ 
$X_2 Z_3$ \\ $Z_2 X_3$ \\ $X_2 Z_4$ \\ $Z_2 X_4$ \\ 
$X_2 Z_5$ \\ $Z_2 X_5$
\end{tabular} & 
\begin{tabular}[c]{@{}c@{}}
$X_0 Z_9$ \\ $Z_0 X_9$ \\ $X_0 Z_{10}$ \\ $Z_0 X_{10}$ \\ 
$X_0 Z_{11}$ \\ $Z_0 X_{11}$ \\ $X_1 Z_9$ \\ $Z_1 X_9$ \\ 
$X_1 Z_{10}$ \\ $Z_1 X_{10}$ \\ $X_1 Z_{11}$ \\ $Z_1 X_{11}$ \\ 
$X_2 Z_9$ \\ $Z_2 X_9$ \\ $X_2 Z_{10}$ \\ $Z_2 X_{10}$ \\ 
$X_2 Z_{11}$ \\ $Z_2 X_{11}$
\end{tabular} \\ \hline
\end{tabular}
\end{table}

After obtaining a Bell pair from simulation, we compare the byproduct stabilizer to every possible stabilizer generator as shown in table \ref{tab:bell_stabilizer_results}. For example, after simulations with the configuration with two RGSs each with three arms, we compare the stabilizer result to column 3 of the table and counting the match stabilizers as the number of Bell pairs.

\section{Analysis of Connection Strategy}
In this work, we investigate the RGS connections for maximizing the number of Bell pairs created. For a complete bipartite RGS core (all-to-all connection for the inner qubits), it is known that a complete bipartite graph proposed by Ref.\cite{Azuma2015-rm} allows for the extraction of at most one Bell pair. We first randomly remove edges in the middle section, perform the $Z$ measurement at the end nodes, and then measure every node in the middle with the $X$ basis.

In our simulations\footnote{\href{https://github.com/PankidT/StimRGS\_v1}{https://github.com/PankidT/StimRGS\_v1}}, the expected number of extractable Bell pairs is determined by 
\begin{equation}
    \langle n_{\text{bell}} \rangle = \frac{\sum_{i=0}^{n_{\max}} (f_i \cdot n_i)}{N},
    \label{eq:expected_bellpair}
\end{equation}
where the summation in the numerator represents the total number of Bell pairs observed throughout the entire experiment, calculated by multiplying each specific number of Bell pairs ($n_i$) extracted from the RGS after the $Z$ and $X$ measurements by its corresponding frequency or occurrence count ($f_i$). Here, 
$i=0$ indicates that no resource was found in that trial, while 
$i=n_{max}$ corresponds to the case where the maximum number of resources was found. This total count is averaged over the total number $N$ of trials performed.

\subsection{Expected Number of Bell Pairs}
In the following, we present heatmaps that illustrate the expected number of Bell pairs for different numbers of RGS arms. The number of possible edges among the inner qubits does not exceed $m^2$. 

\begin{figure*}
    \center
    \includegraphics[width=0.9\linewidth]{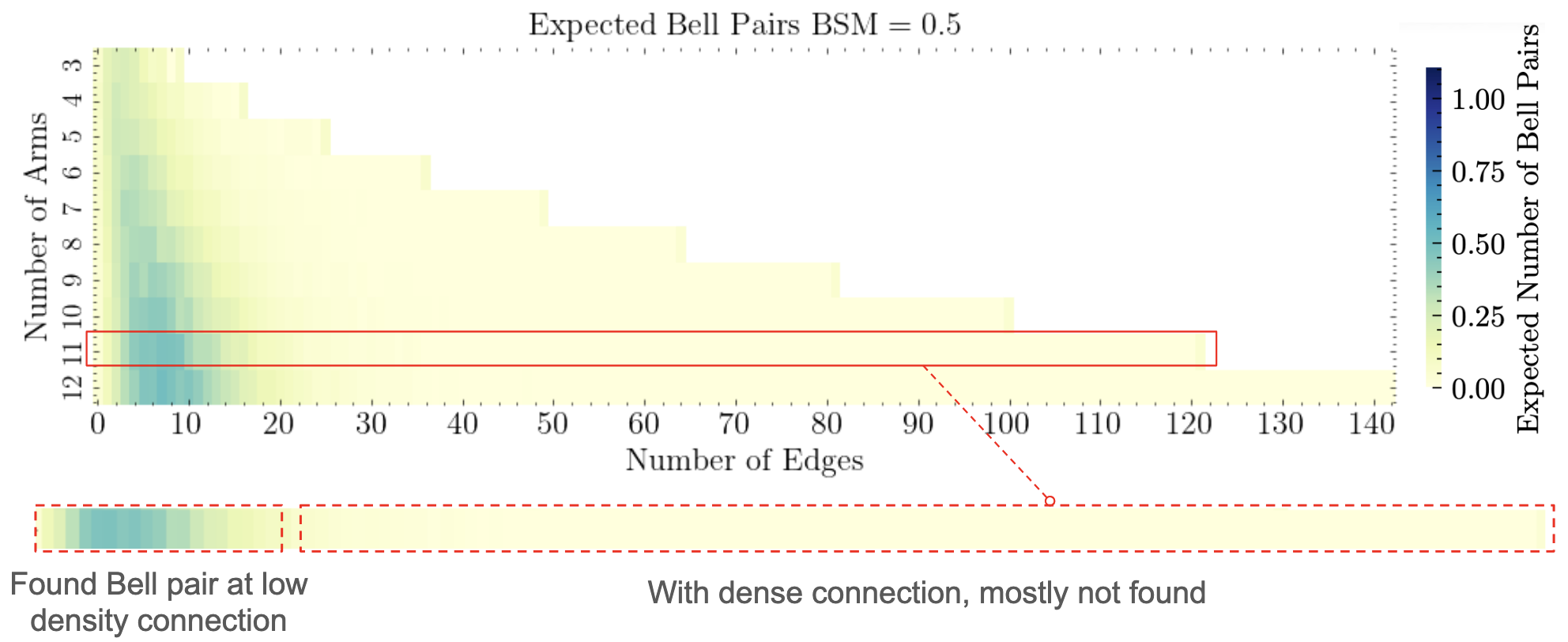}
    \caption{Heat map showing the expected Bell pair yield for various RGS configurations. The top and bottom plots correspond to Bell-state measurement (BSM) success rates}
    \label{fig:heat_maps}
\end{figure*}

The expected Bell pair yield is generally maximal at low to moderate edge densities, as shown in Fig. \ref{fig:heat_maps}. Specifically, the highest yields (dark blue regions) are observed when the number of edges is significantly reduced from that of a complete bipartite RGS, suggesting that sparse or judiciously structured networks can be highly efficient for entanglement generation. As the number of edges increases towards that of a complete bipartite graph, a gradual reduction in the expected Bell pair yield is observed. This decrease suggests that over-connectivity, potentially due to increasing the number of possible non-optimal paths or introducing excessive noise/decoherence channels, negatively impacts the overall efficiency of entanglement distribution.

The results suggest an inverse relationship between network complexity (high edge density) and entanglement generation efficiency. The highest expected Bell pair yields are achieved in sparse network topologies and are critically dependent on high BSM fidelity. This finding has significant implications for the optimal design and allocation of resources in practical quantum communication networks.

\subsection{Analysis of Edge-Bell Pairs}
In this section, we present two simulation results that aim to identify the optimal method to maximize the number of Bell pairs in a given RGS structure.

\begin{figure}[h!]
    \centering
    \includegraphics[width=0.45\textwidth]{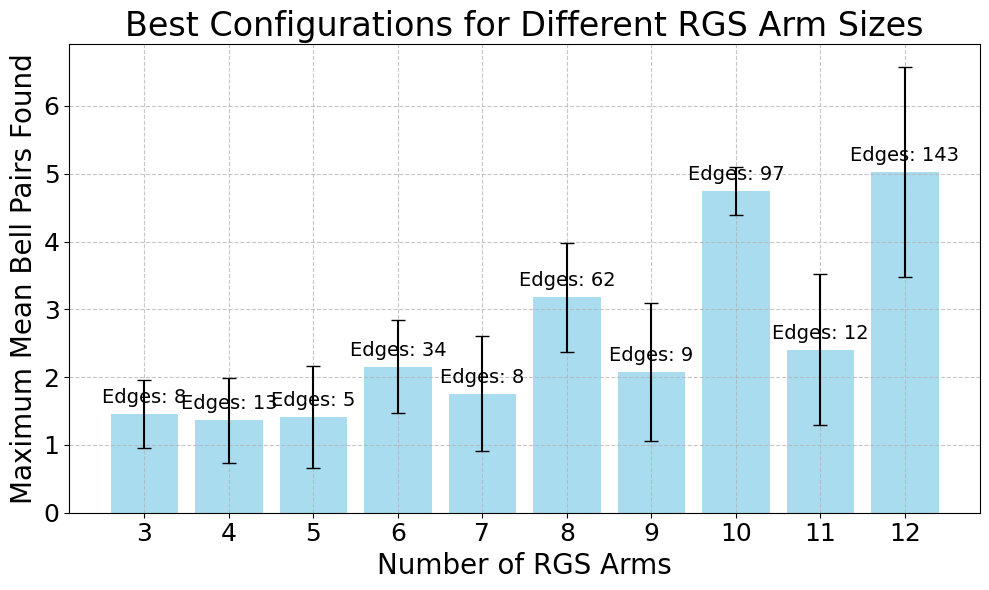}
    \caption{Maximum mean Bell pair found for various RGS arm sizes. The number of optimal number of edges for each arm size is indicated above each bar.}
    \label{fig:rgs_best_configuration}
\end{figure}

First, the maximum mean Bell pair yield achieved for various RGS arm sizes, along with the corresponding optimal number of edges required to achieve that yield is shown in Fig. \ref{fig:rgs_best_configuration}. It demonstrates how the most efficient network topology (in terms of edge count) changes as a function of the number of RGS arms.

Secondly, we investigate the impact of edge density on entanglement generation by fixing a network scale at 6 RGS arms while varying the number of edges between rows. Figure $\ref{fig:heat_maps}$ shows the resulting mean number of Bell pairs found as a function of the edge count, determining the optimal edge density for this specific arm size.

From both results, it can be observed that the number of arms significantly affects the number of extractable Bell pairs. In most cases, reducing the number of edge connections also increases the expected number of Bell pairs.

\subsection{Analysis of Multiple Hops}
This section presents the results exhibiting how different network hop counts influence the number of edges in the RGS. Investigation focuses on the relationship between network hop variations and the resulting edge count of the RGS through simulation analysis.

\begin{figure}[h!]
    \centering
    \includegraphics[width=0.48\textwidth]{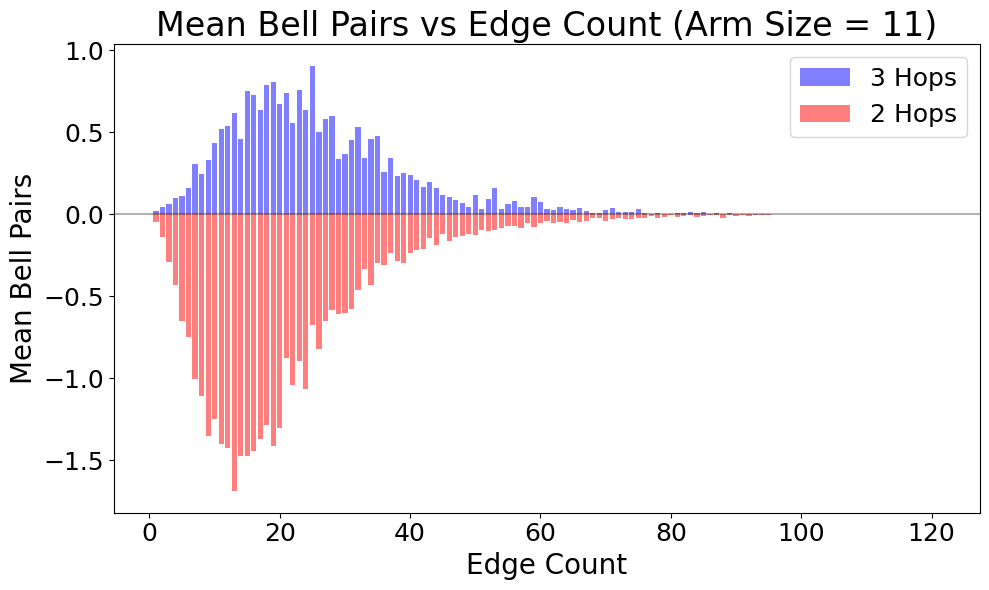}
    \caption{Comparison of the mean number of Bell pairs as a function of edge count for RGS 11 arms ($m = 11$). Blue and red bars represent the results for three-hop and two-hop communication, respectively. The plots show that as the number of hops increases, the edge count corresponding to the maximum mean Bell pair shifts toward higher values.}
    \label{fig:different_hops_analysis}
\end{figure}

The results indicate that, in order to establish communication lines with an increasing number of hops, the number of edges within the inner RGS should be correspondingly increased. As illustrated in Fig. \ref{fig:different_hops_analysis}, a comparison of the expected Bell pair analysis between the two-hops and three-hops situations shows that the number of edges corresponding to the maximum mean Bell pair tends to shift toward higher values in the three-hops case. Furthermore, when varying the number of RGS arms, this overall behavior remains consistent.

\begin{figure}[h!]
    \centering
    \includegraphics[width=0.4\textwidth]{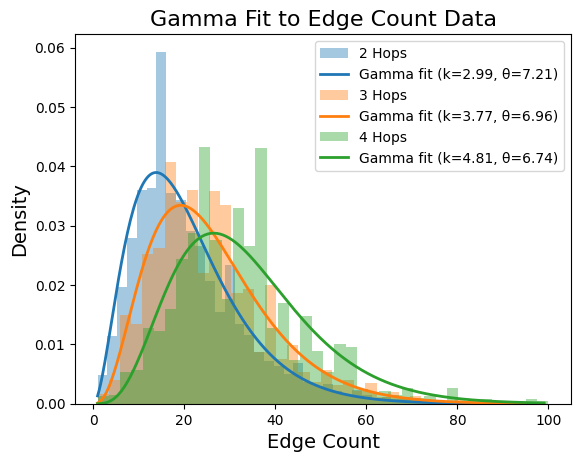}
    \caption{Density histogram of the simulated Edge Counts for 2, 3, and 4 Hops (Arm Size = 12) in the RGS. The empirical distributions are modeled by the Gamma probability distribution. The fitted parameters show that as the Hop Count increases, the shape parameter ($k$) increasing with $k=2.99$, $k=3.77$, and $k=4.81$. Concurrently, the scale parameter ($\theta$) ranging from $\theta=7.21$, $\theta=6.96$, and $\theta=6.74$, suggesting a consistent rate of edge generation regardless of the path length.}
    \label{fig:gamma_fit}
\end{figure}

To provide more insights, we also present the results for three different hop configurations in Fig. \ref{fig:gamma_fit} and table \ref{tab:gamma_analysis_table}. The graph clearly shows that as the number of hops increases, the number of edges required for effective communication increases correspondingly.

From Fig. \ref{fig:gamma_fit}, the empirical distributions exhibit pronounced right-skewness, indicating that a skewed probability model is necessary for accurate analytical characterization. Because the edge count is a positive-valued random variable that effectively results from the sum of multiple independent waiting-time processes, the Gamma distribution is a natural choice for modeling this behavior, which for definiteness, we express it as
\begin{equation}
f(x;k,\theta)=\frac{1}{\Gamma(k)\theta^k}x^{k-1}e^{-x/\theta}, \,\,\,\, x>0,
\end{equation}
where $x$ denotes the edge count, $k$ the shape parameter, and $\theta$ the scaling parameter (or the inverse of rate of events).

\begin{table}[]
\centering
\caption{The Gamma parameters can be directly related to the distribution's theoretical Mean ($\mu$) and Variance ($\sigma^2$).}
\label{tab:gamma_analysis_table}
\begin{tabular}{|c|c|c|c|c|}
\hline
\multicolumn{1}{|l|}{Hop Count} & \multicolumn{1}{l|}{$k$ (Shape)} & \multicolumn{1}{l|}{$\theta$ (Scale)} & \multicolumn{1}{l|}{\begin{tabular}[c]{@{}l@{}}Mean \\ $\mu = k\theta$\end{tabular}} & \multicolumn{1}{l|}{\begin{tabular}[c]{@{}l@{}}Variance \\ $\sigma^2 = k\theta^2$\end{tabular}} \\ \hline
2 Hops                          & 2.99                             & 7.21                                  & $\approx$ 21.56                                                             & $\approx$ 155.44                                                                       \\
3 Hops                          & 3.77                             & 6.96                                  & $\approx$ 26.23                                                             & $\approx$ 182.59                                                                       \\
4 Hops                          & 4.81                             & 6.74                                  & $\approx$ 32.42                                                             & $\approx$ 218.52                                                                       \\ \hline
\end{tabular}
\end{table}

\section{Discussion and Conclusion}
RGS is one architecture of quantum repeaters, based on the concept of the all-photonic quantum repeater first proposed by Azuma et al. \cite{Azuma2015-rm}. By utilizing indirect measurements, this approach eliminates the need for physical quantum memories and provides enhanced robustness against photon loss. In past research, some RGS performance \cite{PhysRevLett.134.190801, zhan2023performance, Hilaire_2021} have analyzed on the aspects of success probability and error analysis. In this work, we examine the RGS topology via the connection of its qubits aimed at maximizing the number of extracting Bell pairs. 

Our results illustrate the dependence of the expected Bell-pair yield on the number of edges and the number of arms within the selected network configurations. On the impact of edge density, a clear trend emerges regarding the relationship between network connectivity (quantified by the number of edges) and the yield of entangled pairs. For performance at low edge density, the expected Bell-pair yield is generally maximal at low to moderate edge densities. Specifically, the highest yields are observed when the number of edges is significantly reduced from that of a complete bipartite graph, suggesting that sparse or judiciously structured networks can be highly efficient for entanglement generation. On the other hand, performance is degraded at high edge density. As the number of edges increases towards that of a complete bipartite graph, a gradual reduction in the expected Bell-pair yield is observed. This decrease suggests that over-connectivity, potentially due to increasing the number of possible non-optimal paths or introducing excessive noise/decoherence channels, negatively impacts the overall efficiency of entanglement distribution. As the number of network hops increases, the RGS must be equipped with higher inner-qubit connectivity to maintain a sufficient yield of extractable Bell pairs for entanglement swapping across the hops. Thus, the expected resource requirement shifts toward using RGSs with higher inner-qubit connectivity.

\subsubsection*{Future Outlook} We emphasize that  in this work we consider the problem of how many Bell pairs can be extracted from the graph states, given a measurement basis. Many open questions remain for future research centering around how measurement bases are selected. How should one select the basis at a node? Would this basis selection depend on the success of BSM? Is there a need for two-way classical communication to select the basis? See, for example, Ref.\cite{Azuma2015-rm}. Furthermore, we speculate that delay lines or quantum memories should enhance the efficiency of entanglement extraction, but how to optimally incorporate them with RGS remains open.

\ifCLASSOPTIONcompsoc  
  \section*{Acknowledgments}
  This project was supported by the JST Moonshot R\&D program under Grant No. JPMJMS226C and by the Faculty of Science, Mahidol University. P. Chianvichai is grateful to the Development and Promotion of Science and Technology Talents (DPST) project for the scholarship to study at Mahidol University and internship at Keio University. P. Chianvichai acknowledges the use of Gemini for English corrections in drafting and revising the manuscript.
\else
\section*{Acknowledgment}

\fi

\bibliography{qcnc-reference}
\bibliographystyle{IEEEtran}

\end{document}